\documentstyle[11pt,aspconf]{article}
\input tex.def
\input psfig.tex
\begin{document}
\title{Low-Luminosity Seyfert Nuclei\altaffilmark{1}}

\author{Luis\,C.\ Ho\altaffilmark{2}, Alexei\,V.\ Filippenko\altaffilmark{3}, 
and Wallace\,L.\,W.\ Sargent\altaffilmark{4}} 

\altaffiltext{1}{To appear in {\it IAU Colloq. 159, Emission Lines in Active 
Galaxies: New Methods and Techniques}, ed. B.~M. Peterson, F.-Z. Cheng, \& 
A.~S. Wilson (San Francisco: ASP), in press.}

\altaffiltext{2}{Harvard-Smithsonian Center for Astrophysics, 60 Garden St., 
Cambridge, MA 02138, U.S.A.}

\altaffiltext{3}{Astronomy Department, University of California, Berkeley, 
CA 94720-3411, U.S.A.}

\altaffiltext{4}{Palomar Observatory, 105-24 Caltech, Pasadena, CA 91125, 
U.S.A.}

\begin{abstract}
We describe a new sample of Seyfert nuclei discovered during the course of an 
optical spectroscopic survey of nearby galaxies.  The majority of the 
objects, many recognized for the first time, have luminosities much lower 
than those of classical Seyferts and populate the faint end of the AGN 
luminosity function.  A significant fraction of the nuclei emit broad H\al\ 
emission qualitatively similar to the broad lines seen in classical Seyfert 
1 nuclei and QSOs.
\end{abstract}
\keywords{active galactic nuclei, Seyfert galaxies}

\setcounter{footnote}{1}

\section{Introduction}
Knowledge of the faint end of the luminosity function of active galactic 
nuclei (AGNs) has bearing on a number of issues concerning the evolution of 
this class of objects.  It it unclear at the moment how the luminosity 
of AGNs changes with time, how objects in different luminosity regimes are 
associated with one another, and how galaxies hosting AGNs relate to those that 
do not.  These reasons and others serve as strong motivation for searching 
for low-luminosity AGNs.  

Until recently, however, the data needed to tackle these issues
have largely been either statistically incomplete or of inadequate quality.
While there is general consensus that low-luminosity AGNs
appear to be present in a substantial fraction of nearby galaxies, the 
existing data suffer from a number of shortcomings that make quantitative 
applications difficult (Ho 1996 and references therein).  The situation 
has been vastly improved with the recent completion of an extensive 
optical spectroscopic survey of nearby galaxies using the Hale 5~m telescope 
at Palomar Observatory (Filippenko \& Sargent 1985, 1986; Ho, Filippenko, \& 
Sargent 1995; Ho 1995, 1996).  We acquired long-slit spectra of exceptional 
quality for the nuclear region of a magnitude-limited ($B_T\,\leq$ 
12.5 mag) sample of 486 northern ($\delta\,>$ 0\deg) galaxies; this
yields an excellent representation of ``typical,''  relatively nearby galaxies 
of all morphological types.  The spectra are of moderate resolution 
(full-width at half maximum [FWHM] $\approx$ 100--200 \kms) and cover two 
regions of the optical window (4230--5110 \AA\ and 6210--6860 \AA) containing 
important diagnostic emission lines.  

\begin{figure}
\psfig{file=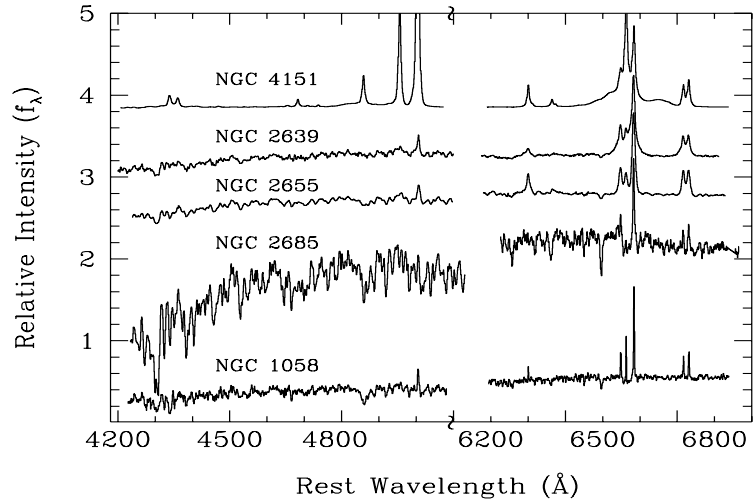,height=2.5truein,width=5.0 truein,angle=-90}
\caption{Sample spectra of Seyfert nuclei from the Palomar survey.  Note the 
diversity of emission-line strengths relative to the stellar component.  
From Ho, Filippenko, \& Sargent (1995).} \label{fig1}
\end{figure}

Here, we take advantage of this unique data set to examine the population of 
Seyfert nuclei contained in it (Fig.~\ref{fig1}
shows a few representative spectra).  Some of the objects have been previously 
cataloged as Seyferts, but many are newly discovered.  We define emission-line 
objects as Seyfert nuclei strictly from the line-intensity ratios of the 
{\it narrow} emission lines (see, e.g., Veilleux \& Osterbrock 1987; Ho, 
Filippenko, \& Sargent 1993; Ho 1996).  Reference will be made to the broad 
emission lines only when further distinction into Seyferts of various 
subclasses (e.g., Seyfert 1, 2, and intermediate types) is called for 
(\S\ 3). The closely related class of low-ionization nuclear emission-line 
regions (LINERs; Heckman 1980) in the sample has been discussed elsewhere 
(Ho 1996) and will not be considered in detail here, except with reference 
to the subset of nuclei in which broad-line emission was detected (\S\ 4).

\section{Statistics of Seyfert Nuclei in Nearby Galaxies}
Objects classified as Seyfert nuclei according to their spectroscopic 
properties comprise 13\% of the survey galaxies ($B_T\,\leq$ 12.5 mag).  The 
subset of 62 objects contains many nuclei of very low luminosity, with a 
median narrow $L_{{\rm H}\alpha}$ = 6\e{38} \lum.  Were it not for the high 
sensitivity of the Palomar survey, many of the objects would either have gone 
unnoticed or been misclassified.  An additional, crucial factor influencing 
the reliability of the classification is the treatment of starlight 
subtraction, as described at length by Ho, Filippenko, \& Sargent (1996).

Our detection rate of Seyfert nuclei is very much higher than that based on 
the Markarian survey ($\sim$1\%; Huchra \& Sargent 1973), and 
still significantly higher than that deduced in later studies which were 
more sensitive to faint nuclei than the Markarian survey (Stauffer 1982; 
Keel 1983; Phillips, Charles, \& Baldwin 1983; Huchra \& Burg 1992; Maiolino 
\& Rieke 1995).  The objects from the Palomar survey, therefore, constitute 
among the lowest luminosity AGNs known.

As was known from previous studies, we find that Seyfert nuclei are 
preferentially hosted 
by galaxies of early Hubble type --- 81\% are found in galaxies of type 
Sbc or earlier.  Contrary to popular belief, Seyfert nuclei do not completely 
shun elliptical galaxies; the detection rate in ellipticals (12\%) is 
essentially identical to that of all disk systems (13\%; S0--Sd).

As discussed in more detail by Ho (1996), several observable parameters depend 
on luminosity when our sample is compared with others containing sources of 
higher luminosity (e.g., Koski 1978).  (1) The most conspicuous difference is 
the tendency for the narrow lines in low-luminosity objects to have smaller 
line widths: the median FWHM of the Seyferts in our sample is $\sim$300 \kms, 
whereas it is conventionally assumed to be larger than this (e.g., Shuder \& 
Osterbrock 1981).  This result reflects the known correlation between line 
luminosity and line width in Seyfert nuclei (Whittle 1992a).  (2) The electron 
density, as traced by the [S~II] \lamb\lamb6716, 6731 doublet, also exhibits 
a noteworthy pattern with luminosity; the density appears to decrease 
systematically with luminosity.  (3) The level of ionization in low-luminosity 
Seyferts typically, but not always, falls 
near the low end of the distribution in Seyferts; the median 
[O~III] \lamb5007/H\bet\ $\approx$ 6 in our sample, close to the 
``low-ionization'' regime of Seyfert 2 and narrow-line radio galaxies (Koski 
1978).  This is consistent with the general absence of He~II \lamb4686 or 
[Fe~X] \lamb6375 in our spectra.  (4) As in other samples (Keel 1980; McLeod 
\& Rieke 1995; Maiolino \& Rieke 1995), the probability of detecting Seyfert 
nuclei is higher for galaxies that are more face-on; however, the inclination 
bias is much less pronounced in our sample, most likely because of the greater 
sensitivity of our survey.

\begin{figure}
\psfig{file=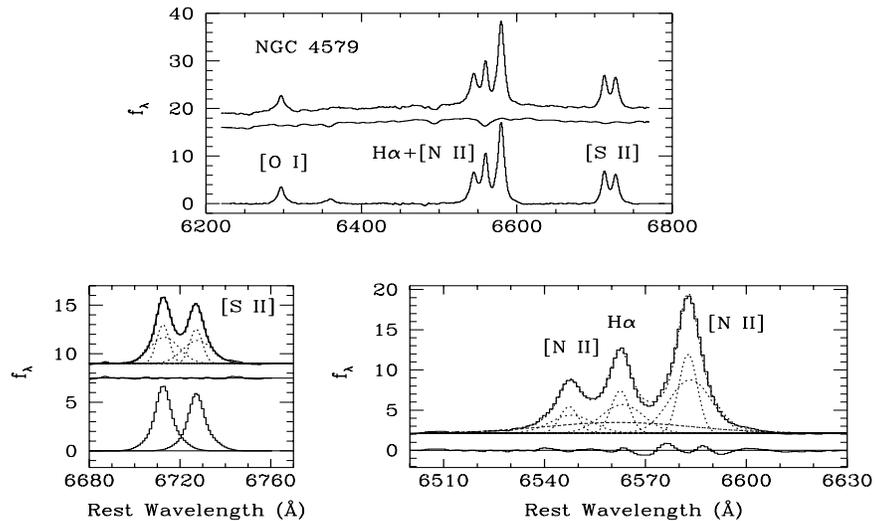,height=2.5truein,width=5.0 truein,angle=-90}
\caption{Example of the technique used to search for broad H\al\ emission
in NGC~4579.  The {\it top} panel shows the removal of
starlight in order to obtain a pure emission-line spectrum.  An empirical
narrow-line profile is generated from the [S~II] \lamb\lamb6716, 6731 lines
({\it bottom left}), which is then used to constrain the narrow components
of H\al\ and [N~II] \lamb\lamb6548, 6583 ({\it bottom right}). A broad
component of H\al, represented as a Gaussian with FWHM = 2300 \kms, is needed
to achieve a satisfactory fit.} \label{fig2}
\end{figure}

\section{Subtypes of Seyfert Nuclei}
It has long been known that the relative strength of the 
broad-line component in Seyfert nuclei exhibits a continuous gradation 
ranging from near complete dominance of the total flux of the permitted 
lines (Seyfert 1s) to its virtual absence (Seyfert 2s).  Osterbrock (1981) 
devised the so-called intermediate types (1.2, 1.5, 1.8, 1.9) to 
empirically categorize this diversity.  Because it becomes increasingly 
challenging to measure a very weak broad-line component in low-luminosity 
AGNs, even H\al\ and H\bet, previous studies of intermediate-type
Seyferts have concentrated mainly on fairly bright nuclei.

Our new survey is well suited for searching for weak broad-line emission, 
particularly that associated with the H\al\ line --- indeed, the survey was 
initially designed largely with this goal in mind (Filippenko 
\& Sargent 1985).  Analysis of the starlight-subtracted spectra (Ho \etal 
1996) reveals that faint broad H\al\ emission is detected in 33 nuclei, and 
with less certainty in another 16.  Thus, roughly 10\% (49/486) of 
all nearby, bright galaxies, contain an AGN with a visible broad-line region.  
If we adhere to the classification scheme based on the narrow-line spectrum, 
half of the sources strictly qualify as LINERs (i.e., because of their 
low [O~III]/H\bet), but clearly all of the objects must be physically related.

The broad H\al\ emission measured typically has FWHM $\approx$ 2200 \kms, 
similar to that seen in classical broad-line AGNs, but the luminosity is
much lower: the median broad H\al\ luminosity is $\sim$10$^{39}$ \lum.  Since 
the broad H\al\ component usually comprises only a small portion of the 
H\al+[N~II] \lamb\lamb6548, 6583 blend, and in general emission lines 
from the nucleus contribute but a minor fraction of the total integrated light
in the spectrum, careful attention to starlight subtraction and line 
decomposition is vital to the detection of this elusive feature.  
Figure 2 gives an illustration of the steps involved in the analysis.

Among the 24 Seyferts in the broad-H\al\ sample, the numerical breakdown into 
the subtypes 1:1.5:1.9 = 1:10:13 (following the convention of Whittle 
1992b).  We did not bother to differentiate between subtypes 1.2 and 1.5, or 
between 1.8 and 1.9, because (1) they can vary from one to the other, (2) 
starlight subtraction often leaves considerable noise in the H\bet\ region, 
and (3) the classification depends on aperture size since the narrow-line 
region is spatially extended.  If we lump all Seyferts 
with any degree of broad H\al\ as type 1, then the ratio of type 1 to 
type 2 is 1 to 1.6.  Similarly, if we adopt a parallel classification 
scheme for the 25 LINERs with broad H\al\ emission, all the 
objects would be of ``type 1.9'' (i.e., only broad H\al\ present, no broad 
H\bet), and the ratio of ``LINER 1s'' to ``LINER 2s'' is 1 to 6.4, where 
we have combined ``pure LINERs'' and ``transition objects'' (see Ho 1996) into 
one category.  If we restrict the comparison to ``pure LINERs'' alone, 
then the ratio is 1 to 4.2.  [Note that the relative numbers of LINER 1s 
and 2s must be interpreted with caution, since it is not clear if all 
narrow-lined LINERs belong to a homogeneous class (e.g., Filippenko 1993).]


\section{Summary}
An extensive optical spectroscopic survey of the nuclei of nearby, bright 
galaxies has recently been completed.  A large, 
well-defined catalog of 62 Seyfert nuclei has been identified.
The results pertaining to this sample of AGNs include the following:
(1) The fraction of galaxies hosting Seyfert nuclei (13\%) is much higher 
than previously thought.  
(2) Most Seyfert nuclei are hosted by galaxies of early Hubble types (Sbc or 
earlier), although they are not restricted solely to spirals.
(3) The emission-line luminosities of the overall sample are much lower than 
those of previously known Seyferts.  Several observed properties, 
including the line width, electron density, ionization level, and inclination 
angle of the host galaxy, seem to change systematically with luminosity.
(4) Weak, broad H\al\ emission has been detected in about 40\% of the 
objects.


\end{document}